\documentstyle[12pt,twoside,fleqn,espcrc1,epsfig]{article}

\newcommand{\AmS}{{\protect\the\textfont2
  A\kern-.1667em\lower.5ex\hbox{M}\kern-.125emS}}
\newcommand{\ave}[1]{\langle {#1} \rangle}
\newcommand{\ket}[1]{| {#1} \rangle}
\newcommand{\bra}[1]{\langle  {#1} |}
\title{Symmetry Conserving Dynamical Mappings}

\author{Zoheir Aouissat
\address{
Institut f\"{u}r Kernphysik, Technische Universit\"at 
Darmstadt, Schlo{\ss}gartenstra{\ss}e 9,\\ D-64289 Darmstadt, Germany}}

\begin{document}
\maketitle

\begin{abstract}
Using the concept of dynamical mappings,
two symmetry conserving nonperturbative approaches are presented. 
The first  is based on the $1/N$-expansion
and sorted out using Holstein-Primakoff mapping. 
The second consists of dynamically mapping the canonical fields into
the corresponding
currents.  
It is argued, 
either by comparing the Fock spaces or the  observables,
that the latter constitutes a higher  approach which 
transcends the $1/N$-expansion and contains the dynamics generated by 
the Gaussian functional approach. 
\end{abstract}

\section{INTRODUCTION}

Solving a theory for interacting fields is one of the major
challenges in quantum field theory (QFT). It is in fact amazing to 
realize that, besides the coupling constant perturbation (CCP),  the  
 analytical (approximate) solutions at our disposal 
are of a semi-classical nature. These are based on expansions in
the  number of colors $(N_c)$, flavors $(N_f)$,
or simply charges $(N)$,  depending on the problem at hand.
I will refer to these generically by the $1/N$-expansion.
The symmetries are known to be preserved by these approaches 
since the dynamics, like in the case of the CCP, 
is sorted out according to an expansion in an arbitrary 
parameter.  
It is well known that the usual CCP
is supported by a Fock space made  of an "uncorrelated"  vacuum
$\ket{0}$ and excited states $\ket{\nu}$ build by the action of 
creation operators of the quantized canonical fields.
Going to higer order in the loops one builds 
correlations perturbatively which leads to a gradual
redefinition of the vaccum at each order.   
An interesting question to raise is: can one envisage
a similar construction for the Fock space in the case of a 
nonperturbative approach e.g. the semi-classical ones?   
In the following I will explore such a 
possibility and show that the idea is fertile and can  even help  
guessing a promising "new" nonperturbative approach. 
The latter, in contrast to the 
semi-classical one,  is an approximate solution  
with a full quantum character.
It transcends the $1/N$-expansion and contains the
Gaussian functional approach (GFA). 
In fact, it was the need for such kind of  solutions of QFT 
which triggered the interest into the GFA. Unfortunately,
it was quickly realized that the latter, being order mixing, is not in general 
 able to treat  the symmetries correctly \cite{GAC}.     
Here I would like to argue that the second nonperturbative symmetry conserving 
approximation, next to the $1/N$-expansion, needs in fact much more 
vaccum-correlations than what the GFA offers. We will see that 
these correlations are of a RPA-type, selected carefully by dynamically mapping
the  canonical fields into the  currents with the corresponding quantum numbers.
The idea  of substituting the currents for the canonical fields is 
in fact not new. 
It was used in the  late 
sixties by Callan, Dashen, Sharp, Sommerfield,  and Sugawara \cite{HIRO}
in an attempt to build a QFT with currents as dynamical
variables giving up the concept of describing  the fields 
with canonical variables.
In the following, I would like to draw a slightly different picture. Although
I will substitute for the asymptotic fields  
the corresponding currents, I will not renounce the use of the 
canonical fields as building blocks of QFT.  
Mapping these canonical fields into the currents will then help in gathering 
the dynamics which dress the asymptotic fields while preserving the symmetry. 
In section 3, we will see how this can be put to work 
in building a nonperturbative pion  by mapping its canonical field 
into the axial current.  
However, I will first, in section 2, revisit the  semi-classical
$1/N$-expansion
approach using the very concept of dynamical mappings.\\
 As an example, consider the toy model of QFT,
the $\Phi^4$-theory with
a continuous $O(N+1)$ symmetry.        
The lagrangian density with the appropriate scaling reads
\begin{equation}
 {\cal L}  =  \frac{1}{2}\left[ \left(\partial_{\mu}{\vec \pi}\right)^2
 + \left(\partial_{\mu}{ \sigma}\right)^2 \right] 
 - \frac{\mu^2}{2} \left[ {\vec{\pi}}^2 + { \sigma}^2 \right]
 - \frac{\lambda}{4N}\left[ {\vec{\pi}}^2 + { \sigma}^2 \right]^2
   + \sqrt{N}c {\sigma}~,
\label{eq1}
\end{equation}
where ${\vec \pi}(x)$ stands for a $N$-components pion field  and
$\sigma(x)$  its chiral partner. 

\section{HOLSTEIN-PRIMAKOFF MAPPING}
First let us see how the concept of symmetry conserving dynamical mapping 
(SCDM) can be  used to retrieve the well known $1/N$-expansion.
It is clear that, due to the Bose statistics,
the pion wave function induces direct and exchange contributions 
(Hartree and Fock terms) which are of two distinct orders in the $1/N$
counting.
Therefore to hinder any order mixing in the $1/N$-expansion one should,
whatever the  procedure used, only allow  the Hartree terms
as leading contributions  and relegate the Fock terms to the sub-leading
orders. This is, however,  only possible if the pion wave-function is 
severely truncated, leading to a particle which doesn't fully enjoy the quantum
statistics or, in other words, to a Hartree particle.  
Sorting out the dynamics according to this scheme can be achieved 
by means of a pion-pair bosonization via the so-called
Holstein-Primakoff mapping (HPM). The latter appeared first in the early 
fourties \cite{hopr} as a realization
of the $SU(2)$ algebra for quasi-spins. It was forgotten ever since 
and reappeared  in the sixties
in the nuclear many-body problem where it was  
used for bosonizing fermion-pairs (see \cite{M93} for a review).
 In the present case 
the HPM for pion-pairs reads (see \cite{ASW} for  details):
\begin{equation}
{\vec{a}}^{+}_{q}{\vec{a}}^{+}_{p} \, \rightarrow \, 
\left( A^+\sqrt{ N \,+\, A^+A } \right)_{q,p}~,\quad\quad
{\vec{a}}_{q}{\vec{a}}_{p} \, \rightarrow \,
\left( {\vec{a}}^{+}_{p}{\vec{a}}^{+}_{q} \right)^+~, \quad\quad
{\vec{a}}^{+}_{q}{\vec{a}}_{p}  \, \rightarrow \, \left( A^+ A \right)_{q,p}~,   
\label{eq2}
\end{equation}
where ${\vec a}^+_q, {\vec a}_q$ stand for the  pion creation and
annihilation operators, while $A_{q,p}$ and $A^+_{q,p}$ 
are real boson operators obeying the Heisenberg-Weyl algebra. This mapping is
made in such a way that the original algebra, obeyed by the pairs
of operators at the  l.h.s of eq.(\ref{eq2}), is also realized by the
ansatz at the r.h.s. The square root is 
to be understood as a formal power series in the operators. Thus the
Hamiltonian of the vector model derived from eq.(\ref{eq1}) will naturally 
inherit a formal expansion of the form:~ 
$H = {\cal H}^{(0)} + {\cal H}^{(1)}+ {\cal H}^{(2)}+ {\cal H}^{(3)}+
{\cal H}^{(4)} +..$, where the superscripts indicate the powers of the
operators $A$ and $A^+$ and also $b$ and $b^+$ for the sigma field. 
At this stage, the content in $N$ of each     
order ${\cal H}^{(p)}$ is not yet specified. Also the formal expansion is
in reality  not unique since the operators are not in normal order. 
Therefore a 
definition of a vacuum is mandatory if one wishes to make any use of the above
expansion. By defining a vacuum for the $A$ and $b$ operators one
 makes 
the HPM dynamical. 
To meet the desired $1/N$-expansion approach this step is indeed  
decisive.    
In other words not any vacuum and thus not any Fock space is able to support this 
approach. It was shown in \cite{ASW} that  the
vacuum of the $1/N$-expansion is a coherent state 
\begin{equation}
| \psi> = exp\left[ \sum_{q} d_{\pi\pi}(q) \,A^+_{q,-q} \,\, + \,\, 
 \ave{b_0}\, b_{0}^+ \right] |0>~,
\end{equation}
which can accommodate condensates of 
the sigma field, denoted here by $\ave{b_0}$, as well as  pairs of Hartree 
pions, denoted by $d_{\pi\pi}(q)$. Assuming this, the Hamiltonian displays then a parallel (and unambiguous)
expansion in the powers of $N$, such that 
\begin{equation}
H=NH^{(0)}+\sqrt{N}H^{(1)}+H^{(2)}+\frac{1}{\sqrt{N}}H^{(3)}
+\frac{1}{N}H^{(4)}+...
\end{equation}
Here the terms $H^{(p)}$ have no content in $N$. 
To gather the dynamics, one has to  diagonalise $H$. Odd powers in
 $\sqrt{N}$ are completely off-diagonal and therefore ought to disappear.
The leading order dynamics is contained in the three lowest terms, thus we
disregard here all higher terms. The term $H^{(1)}$ is washed out by performing
a variational Hartree-Bogoliubov (HB) calculation, using $\ave{b_0}$ and   
$d_{\pi\pi}(q)$ as variational parameters in minimizing the ground state
energy. The bilinear $H^{(2)}$
 is diagonalized by applying a canonical Bogoliubov rotation \cite{ASW} which 
 mixes the operators $b$, $b^+$ and $A$, $A^+$ such that:
\begin{equation}
Q^+_{\vec p}\,=\, X_{\vec p}\, b^+_{{\vec p}}\,
-\, Y_{\vec p}\, b_{-{\vec p}}
\,+\, \sum_{q} \left[ U_{{\vec q},{\vec p}}\, 
A^+_{{\vec q},{\vec p}-{\vec q}}
\,-\, V_{{\vec q},{\vec p}}\, A_{-{\vec q},-{\vec p}+ {\vec q}}\right].
\label{eq328}
\end{equation} 
This is nothing but a $\pi\pi$ RPA-scattering equation 
 coupled to a Dyson equation for the sigma mode.
The vacuum of the theory is  accordingly modified. The latter, 
denoted by $\ket{RPA}$, is implicitly defined by $Q_{{\vec p}}\,\ket{RPA} =0$ 
and explicitly obtained via a unitary transformation\footnote{ 
This transformations is constructed as  a product of three unitary inequivalent
transformations. The first is a unitary squeezing transformation
in the $(b, b^+)$-sector, the second is a similar one in the $(A, A^+)$-sector
and the third is a unitary transformation which mixes both sectors \cite{BHV}.}
 of the coherent state:~ $\ket{RPA} = U_{unitary} \ket{\psi}$.\\
This exhausts the leading order dynamics. 
In a cutoff theory, the $\ket{RPA}$-vacuum, obtained so far, possesses a broken 
phase with a finite sigma-condensate $(\ave{\sigma}\neq 0)$  and two
curvatures; one is the Goldstone boson mass , obtained in the HB mean-field, 
and the second is the sigma mass, obtained in the RPA.    
These are given by 
\begin{eqnarray}
  m_{\pi}^2 &=& \mu^2 + \lambda \left[ I_{\pi} + \ave{\sigma}^2\right]~,
 \quad\quad\quad
  \frac{c}{\ave{\sigma}} = \mu^{2} + \lambda \left[ I_{\pi} + 
  \ave{\sigma}^2 \right]~, \\
m_{\sigma}^2 \,&=&\, \mu^2 + \lambda \left[ I_{\pi} + 3 \ave{\sigma}^2\right]
\,+\, \frac{2 \lambda^4 \ave{\sigma}^2\,{\Sigma}_{\pi\pi}(m_{\sigma}^2)}
{ 1\,-\,  \lambda^2 {\Sigma}_{\pi\pi}(m_{\sigma}^2)}~.
 \label{eq513}
\end{eqnarray}
Here $I_{\pi}$ is the tadpole of the  HB-pion and 
${\Sigma}_{\pi\pi}(p^2)$ stands for the convoluted two HB-pion propagator
(RPA bubble of HB-pions). Naturally, this approach preserves the 
whole hierarchy of Ward identities. 
In particular, the lowest one which expresses
the current conservation (in PCAC sense), $D_{\pi}^{-1}(0) =
\frac{c}{\ave{\sigma}}$, holds. 
Figures 1.a and 1.b show the summed class of diagrams.

\parbox{5.1cm}{{\bf Fig. 1.a.}~~~~BCS solution in a Hartree-Bogoliubov 
(HB) approximation. The pion mass 
and the condensate are given by  two coupled self-consistent
 equations. There is no dynamical mass generation therefore the pion is a
 Goldstone mode}
\hspace*{1.cm}
\parbox{10.cm}
{\epsfig{file=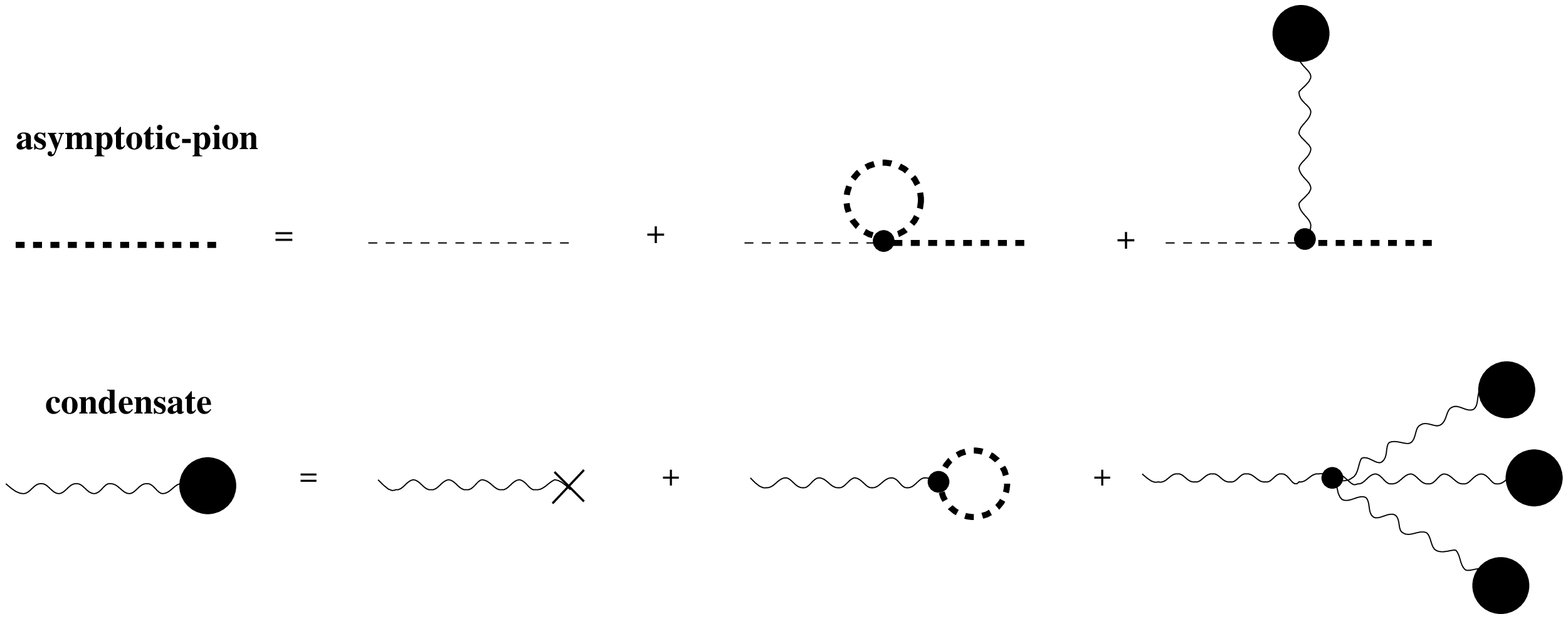,height=10cm,angle=270}}
\vspace*{0.3cm}

{{\bf Fig. 1.b.}~~ Dyson equation for the $\sigma$ mode coupled
 to a RPA equation for $\pi\pi$-scattering. According to this scheme, the sigma
 mass is build perturbatively (in contrast to the self-consistent building of
 the pion and the condensate in Figure 1.a.)}
 
\hspace*{1.0cm}
\parbox{14.cm}
{\epsfig{file=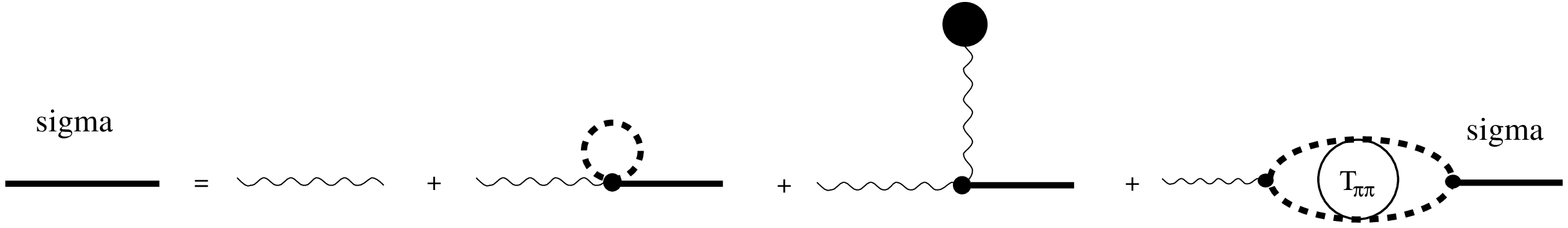,height=10cm,angle=270}}

\section{PIONIC-QRPA MAPPING}
The approach presented in the previous section is very appealing from many
aspects and particularly from its symmetry conserving character. However, 
it has a serious drawback.
The pion, constructed so far, is a Hartree particle thus  a semi-quantum 
(semi-classical) "object".
It is also the building block for all higher n-point functions, as
suggested by the whole hierarchy of Ward identities.  
Attempts made with the GFA failed so far  to correct for this 
\cite{GAC}. 
Indeed, assuming the full wave function of the pion (instead of truncating it)
induces an uncontrollable order mixing which inevitably "destroys the symmetry".   
As stated in the introduction, one way out is to map the canonical pion field 
into the axial current. This idea is supported by  the
exact (Goldstone) statement: ~$Q_5^a \ket{vac} \propto \ket{\pi^a}$~,
which allows to build a pion state by acting 
with the symmetry generator on the full correlated vacuum of
the theory. 
In an effective model, the 
generator $Q_5^a$ is simply given by Noether's theorem. Therefore one can 
use the field structure of $Q_5^a$ to model an excitation operator for 
the asymptotic pion field. In the present case, the  creation 
operator of the iso-vector pion takes the form
\begin{equation}
\vec Q_{\pi}^+ =
 X^{(1)}_{\pi} \vec a^{+}_0 -  Y^{(1)}_{\pi} \vec a_0
 +
 \sum_{q}
  \left[ X_{\pi}^{(2)}(q)  b^+_q \vec a^{+}_{-q}
  -  Y_{\pi}^{(2)}(q)  b_{-q} \vec a_q 
 +
 X_{\pi}^{(3)}(q)  b^+_q \vec a_{q}
  -  Y_{\pi}^{(3)}(q)  b_{q} \vec a^{+}_{-q} \right].~~ 
\label{eq210}
\end{equation}
Here the operators $\vec a_q$ and $b_q$ represent the canonical 
pion- and sigma- fields. The $X$ and $Y$ amplitudes are fixed
dynamically. Using these as variational variables to minimize the 
 the  pion-state energy   
$ 
(\delta \frac{\bra{\pi} H \ket{\pi}}{\langle \pi|\pi \rangle} = 0~)
$
leads to the so-called Rowe equation of motion
\begin{equation}
 \bra{RPA }\left[\delta {\vec Q}_{\pi} ,  
 \left[H , {\vec Q}_{\pi}^+\right]\right]\ket{ RPA}=
m_{\pi}\bra{RPA }\left[ \delta {\vec Q}_{\pi} 
,{\vec Q}_{\pi}^+\right]\ket{ RPA}~,
\label{eq329}
\end{equation} 
where H is the  Hamiltonian of the model, $m_{\pi}$ is the
pion mass (excitation energy to create a pion at rest) and $\ket{RPA}$ is
an approximate ansatz to the full correlated vacuum $\ket{vac}$,
 defined implicitly by :~ ${\vec Q}_{\pi}\ket{RPA} = 0$.  
The eigenvalue problem in eq.(\ref{eq329}), in its present variational form, 
is known as the self-consistent RPA which can not be solved 
in practice. Therefore one uses in general
the quasi-boson assumption which approximates the bilinear 
${\vec Q}_{\pi}$ by a boson. 
Thus eq.(\ref{eq329}) is linearized. In the exact chiral limit $(c=0)$, 
one of its solutions, if successfully normalized, 
  has zero energy $(m_{\pi}=0)$.  The normalization of this Goldstone
 solution can in fact be achieved
   by optimizing the RPA basis. This is done by dynamically  
 mapping the original canonical pion $({\vec a}, {\vec a}^+)$ and
  sigma $(b, b^+)$ fields
  into  Hartree-Fock-Bogoliubov (HFB)
 fields\footnote{This  is in fact the minimal procedure to achieve
 the normalization of the Goldstone solution. Other normalization
 procedures, based on Higher-RPA, and which allow to 
 gather  more dynamics than the present approach, are also possible. 
 Further discussion on this point is  deferred to a coming work.}  
\begin{equation}
  \vec\alpha_q^+ = u_{q}\vec a_q^+ - v_{q}\vec a_{-q}~,\quad\quad\quad
  \beta ^{+}_{q} = x_{q}b^{+}_{q} -
    y_{q}b_{-q} - w^*\,\delta_{q0}~.
\label{eq204}
\end{equation}
Here  $u, v, x, y, w$ are  variational functions chosen to minimize the
energy of the  vacuum of the theory. The latter is given, up to an unimportant factor, by
 the  squeezed state
\begin{equation}
\ket{\Phi} =\exp\left[ \sum_q  \frac{v_q}{2u_q}\vec a_q^+\vec a_{-q}^+ 
\, + \, \frac{y_q}{2x_q} b_q^+ b_{-q}^+\, +\, \frac{w}{2x_0} b_0^+ \right]
\ket{0}~. 
\label{eq335}
\end{equation}
The  dynamics gathered at this HFB mean-field appear in the following
set of equations: 
\begin{eqnarray}
 {\cal E}_{\pi}^2&=& \mu^2 + \lambda \left[\frac{N+2}{N} I_{\pi} + 
 \frac{1}{N}I_{\sigma} +   \ave{\sigma}^2\right]~,
 \quad\quad
 {\cal E}_{\sigma}^2= \mu^2 + \lambda \left[ I_{\pi} + 
  \frac{3}{N} I_{\sigma} + 3  \ave{\sigma}^2\right]~,
\nonumber\\
  \frac{c}{\ave{\sigma}} &=& \mu^{2} +
   \lambda \left[   I_{\pi} + \frac{3}{N} I_{\sigma} +  \ave{\sigma}^2 \right]~. 
 \label{eq16}
 \end{eqnarray}
which consist of three coupled self-consistent BCS gap
equations that give the condensate $(\ave{\sigma})$ and  the two   curvatures 
$({\cal E}_{\pi}, {\cal E}_{\sigma})$ 
of  the squeezed vacuum $\ket{\Phi}$.\\ 
Here  $I_{\pi}$ and $I_{\sigma}$ are, respectively, the tadpoles for the pion and sigma
quasi-particles (with the Hartree and Fock terms considered together). 
This is precisely the dynamics generated by the GFA where 
${\cal E}_{\pi}$ stands for the asymptotic pion mass. This is, however,
 clearly wrong.
Indeed, in the exact chiral limit ($c=0$) and for a
finite condensate, the
curvature ${\cal E}_{\pi}$ does not vanish (see also \cite{GAC}). Therefore the squeezed state and
equally the Gaussian functional can not be regarded as a viable vacuum for the
theory since the Goldstone theorem is violated. 
However, the RPA ground state, 
as implicitly defined by ${\vec Q}_{\pi}\ket{RPA} = 0$, 
is a good candidate for a vacuum with broken symmetry.
The latter, in the case of
the quasi-boson assumption (used here), is explicitly obtained by 
an unitary transformation
of the squeezed state: $\ket{RPA}= U_{unitary} \ket{\Phi}$ \footnote{In the case
of the QBA assumption, the way of building the unitary transformation is
similar to the one sketched in footnote 1. Because of  the infinite
degrees of freedom,  both vacuums are in fact
inequivalent. }.
The curvature along
the valley of this ground-state is given by the RPA eigenvalue $m_{\pi}$ 
and reads: 
\begin{equation}
m_{\pi}^2 \,=\, \frac{c}{\ave{\sigma}} \quad+\quad  \frac{ 2\lambda^2}{N}\,
\frac{  \left[{\cal E}_{\pi}^2 \,-\,
 {\cal E}_{\sigma}^2 \right]
\left[\Sigma_{\pi\sigma}(0)
 \,-\, \Sigma_{\pi\sigma}(m_{\pi}^2)\right]}
{ 1\quad-\quad \frac{2 \lambda^2}{N} \Sigma_{\pi\sigma}(m_{\pi}^2)}
\label{eq212}
\end{equation}
where $\Sigma_{\pi\sigma}(p^2)$ is the convoluted quasi-pion and
quasi-sigma propagators (see figure 2).
It is clear, from eq.(\ref{eq212}), that the asymptotic pion 
is not only  highly nonperturbative in the coupling $\lambda$ 
but also has a non-trivial
content in  $N$, in contrast to the HB-pion of section 2. 
It is, however, still a Goldstone  
mode, since for $c=0$ and  $\ave{\sigma}\neq 0$  a zero pion
mass exists. Furthermore it is easily verified that 
the Ward identity, 
$D_{\pi}^{-1}(0) = \frac{c}{\ave{\sigma}}$, holds here too. 

\vspace*{0.4cm}
\parbox{4.5cm}
{{\bf Fig.2.}  Diagrammatic representation of the  collected dynamics.
In step I, the optimized quasi-particle basis is build  as a BCS  
solution in the HFB-approximation. 
In step II, the quasi-particle states are scattered in a Lippmann-Schwinger
equation. In step III, a mass operator is build out of the full vertex 
$T_{\pi \sigma}$ and inserted in a Dyson equation to generate the asymptotic
Goldstone pion.  }
\hspace*{0.7cm}
\parbox{10.cm}{
\epsfig{file=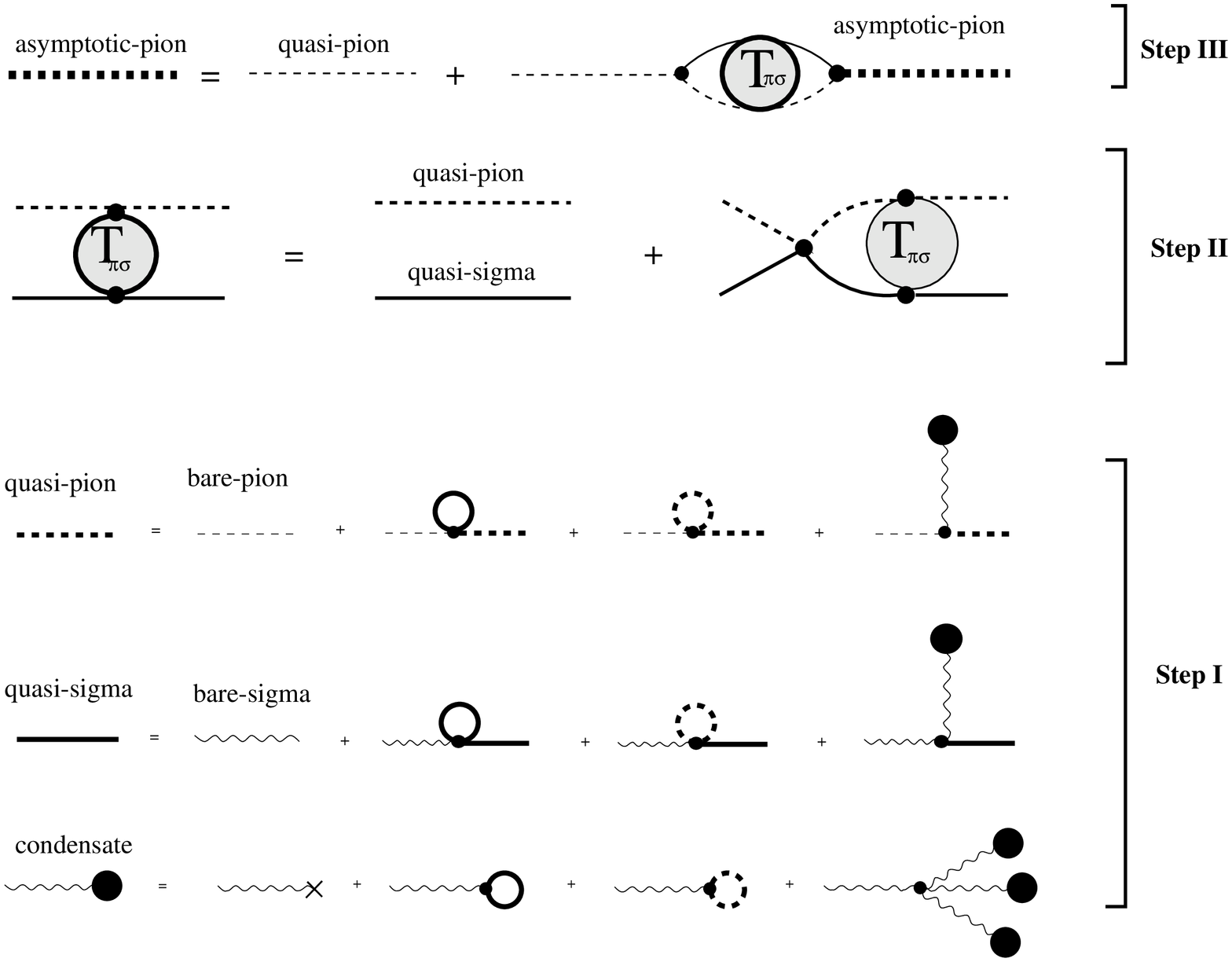,height=11cm,angle=270}}
\section{CONCLUSION}
There is obviously  an urgent need  for developing 
symmetry conserving nonperturbative approaches  with 
tractable analytical solutions to QFT. 
I exposed here the concept of SCDM which  is a
promising tool that gives  a helpful
insight on the structure of the Fock space.   
Besides the HPM which leads to the
$1/N$-expansion, I presented a second  SCDM that relies on a systematic 
 scheme which consists of mapping the 
canonical fields into the corresponding currents. 
The latter mapping was made  dynamical in the  quasi-particle RPA.
 The vacuum of the theory was then found  to have 
more correlations than the vacuums of the $1/N$-expansion and the GFA alike. 
Extensions to higher RPA, finite temperature and baryon density 
as well as to richer dynamics are possible \cite{ABW}. 

\vspace*{0.2cm}
{\bf Acknowledgements}~: I would like to thank G. Chanfray, P. Schuck and J.
Wambach for their interest in this work and for their continuous support.

\end{document}